%% file: main.tex
\documentclass[amsmath,amssymb,aps,prl,twocolumn,showpacs,floatfix,superscriptaddress,nofootinbib]{revtex4-2}
\usepackage[english]{babel}
\usepackage{graphicx}%
\usepackage{dcolumn}%
\usepackage{bm}%
\usepackage{physics}
\usepackage{hyperref}%
\usepackage{color}
\usepackage{ulem} %
\usepackage{multirow}
\usepackage{booktabs}
\usepackage{blkarray}
\usepackage{tabularray}
\usepackage[ruled]{algorithm2e} 
\usepackage[bb=dsserif]{mathalpha}

\definecolor{gr4} {RGB}{34,118,34}

\newcommand{\sx}{\sigma^x}

\newcommand{\sz}{\sigma^z}
\newcommand{\rhoxx}{\rho^{xx}}
\newcommand{\rhozzc}{\rho^{zz}_c}
\newcommand{\vmax}{v_{\text{max}}}

\begin{document}

\title{Recurrence analysis of quantum many-body dynamics}

\author{Tomasz Szo\l{}dra}
\affiliation{Zentrum für Optische Quantentechnologien, Universität Hamburg, Luruper Chaussee 149, 22761 Hamburg, Germany}
\author{Matheus S. Palmero}
\affiliation{Potsdam-Institut für Klimafolgenforschung (PIK), Mitglied der Leibniz-Gemeinschaft, Telegrafenberg A31, 14473 Potsdam, Germany}
\affiliation{Instituto de Ciências Matemáticas e de Computação, Universidade de São Paulo, 13560-970 São Carlos, São Paulo, Brazil}%
\author{Peter Schmelcher}
\affiliation{Zentrum für Optische Quantentechnologien, Universität Hamburg, Luruper Chaussee 149, 22761 Hamburg, Germany}

\date{\today}

\begin{abstract}
Observables of out-of-equilibrium quantum many-body systems display complex temporal behavior that encodes the underlying physical mechanisms but typically resists straightforward interpretations. We introduce recurrence analysis - a nonlinear time-series analysis framework long established for classical dynamical systems - to investigate correlated quantum many-body dynamics. Recurrence plots provide a qualitative fingerprint of simulated or experimental data, while recurrence quantification analysis extracts corresponding numerical descriptors. Applying this framework to quenches from the paramagnetic ground state in the one-dimensional transverse-field Ising model, we observe a clear progression in the recurrence plots of two-site correlations: nearly periodic patterns in the deeply ferromagnetic phase give way to multiscale temporal structures at criticality. Recurrence quantifiers further recover the critical field strength without prior knowledge of the model, establishing recurrence analysis as a versatile tool for characterizing quantum many-body dynamics, including unsupervised detection of quantum phase transitions.
\end{abstract}

\maketitle

Quantum many-body systems represent one of the most active frontiers of modern physics, with questions driving research across fields such as condensed matter, atomic physics, quantum information, and quantum computing. Their non-equilibrium dynamics has attracted interest not only in the context of thermalization \cite{deutsch_quantum_1991,srednicki_chaos_1994}, expected for generic Hamiltonians, but particularly in models with broken ergodicity resulting from phenomena such as integrability \cite{orbach_linear_1958, lieb_absence_1968, pfeuty_ising_1970}, many-body localization \cite{basko_metal_2006,gornyi_interacting_2005,oganesyan_localization_2007,sierant_mbl_2025}, quantum many-body scarring \cite{bernien_probing_2017,turner_weak_2018,serbyn_quantum_2021}, or Hilbert space fragmentation \cite{sala_ergodicity_2020}. In parallel to theoretical studies, advances in ultracold neutral atoms \cite{anderson_observation_1995, lewenstein_ultracold_2012, schaefer_tools_2020}, trapped ions \cite{cirac_quantum_1995,fossfeig_progress_2025}, and superconducting qubit platforms \cite{nakamura_coherent_1999,jiang_advancements_2025} have enabled highly controlled experiments with nearly perfect isolation from the environment, allowing for a direct observation of non-equilibrium unitary quantum evolution \cite{bernien_probing_2017, dborin_simulating_2022, hudomal_ergodicity_2025, liang_observation_2025, fischer_dynamical_2026}.

Unfortunately, even if the observables can be simulated or experimentally measured, their interpretation and characterization in light of complex phenomena occurring in the exponentially large Hilbert space is in many cases far from trivial, especially when the classical limit is not available. For systems in equilibrium, schemes like unsupervised detection of phase transitions with neural networks have been introduced \cite{kottmann_unsupervised_2020, lidiak_unsupervised_2020, kaeming_unsupervised_2021, kottmann_unsupervised_2021, kottmann_variational_2021}. In the time-dependent case the recent developments concentrate on strategies supplementing standard time series visualization and statistical analyses, such as e.g. quantum data sonification \cite{tudoce_sonification_2025,yamada_exploring_2025} or representing many-body wave functions as images \cite{rodriguez_qubism_2012,koczor_fast_2020,barthe_bloch_2023}, both exhibiting an attractive artistic component. Other works employ neural networks \cite{szoldra_detecting_2021} and problem-tailored probes \cite{szoldra_tracking_2023}. While applicable in specific cases, the exploratory and sometimes only qualitative character of these and similar methods adds an extra layer of complexity to the processing of already complex signals.

In contrast, recurrence plots (RPs), originally developed for classical dynamical systems \cite{Eckmann1987}, are a relatively simple, theoretically grounded, and well-understood time series analysis tool \cite{Webber2005,marwan2007recurrence,marwan2008epjst}. They are constructed as two dimensional images by marking all pairs of times at which a trajectory revisits the same neighborhood in an effective phase space within a preset distance threshold $\varepsilon$. The recurrence quantification analysis (RQA) \cite{marwan2007recurrence,MarwanWebber2015} supplements RPs by employing numerical indicators based on the recurrence structures, objectively characterizing the underlying finite-time dynamics and distinguishing periodic, quasi-periodic, chaotic, and stochastic signals. Applications of RPs and RQA range across various scientific domains, such as classical physics \cite{Thiel2002,Zolotova2006,Palmero2023}, physiology and neuroscience \cite{Marwan2002, Acharya2005,Prabhu2020}, climate and earth sciences \cite{spiridonov2017,Trauth2019}, finance and economics \cite{Fabretti2005,Bastos2011}, engineering systems \cite{Nichols2006, Sen2008,kasthuri2019}, and ecological research \cite{ayers2015,zurlini2018,semeraro2020,semeraro2021}. 
Nonetheless, in the context of quantum sciences, the adoption of RPs is rather limited. Previous works concentrate on descriptions of relativistic mean-field dynamics in atomic nuclei \cite{vretenar_nonlinear_1999}, single atoms in electromagnetic fields \cite{sudheesh_recurrence_2010, laha_bifurcations_2020, laha_timeseries_2020, lakshmibala_dynamics_2022}, or deformed quantum harmonic oscillators \cite{Pradeep2020}. While these papers demonstrate that recurrence-based methods can reveal nontrivial structures in quantum-related problems, none of them targets a correlated, truly many-body case.

\begin{figure*}
    \centering
    \includegraphics[width=\linewidth]{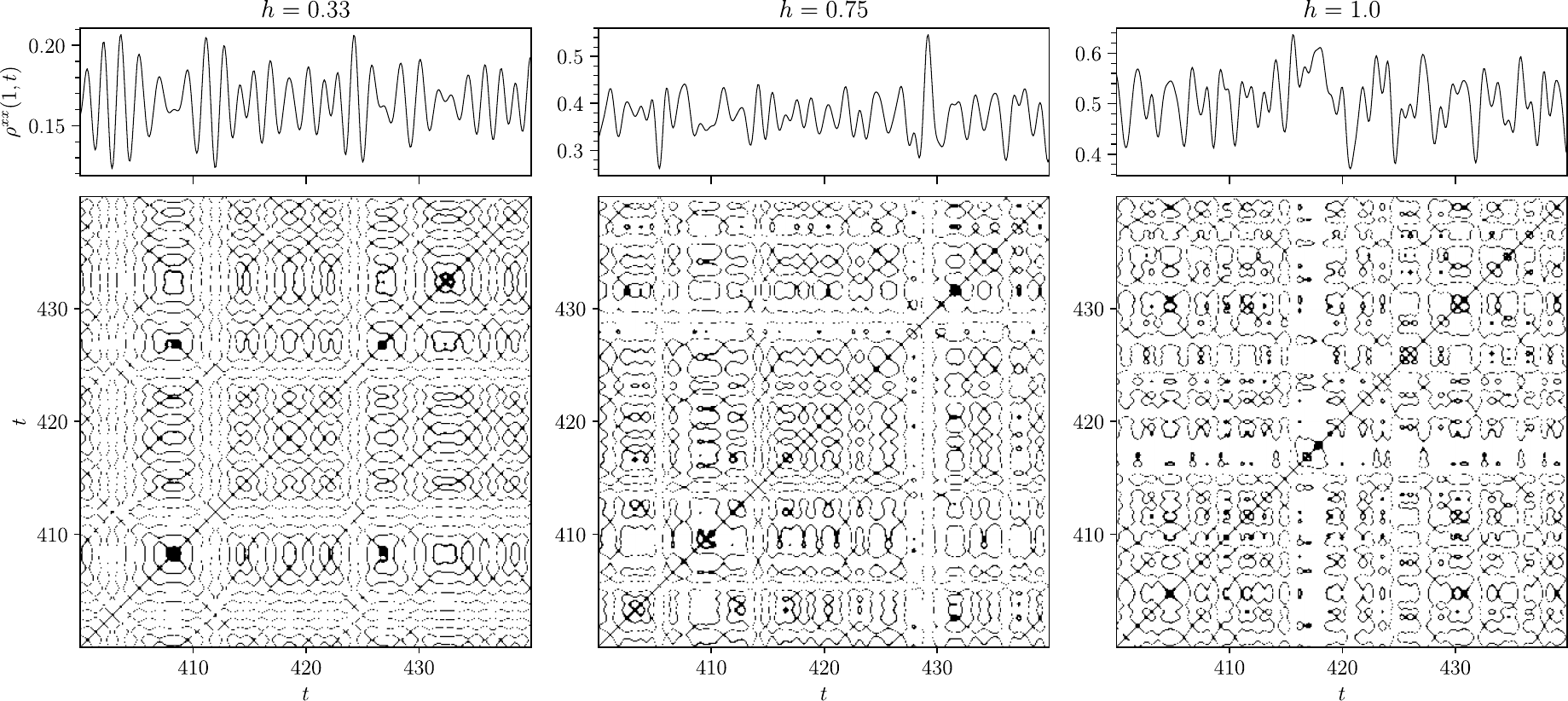}
    \caption{Long-time dynamics of nearest-neighbor correlations in a quench from the paramagnetic state ($h_0 \rightarrow \infty$) to varying $h$ within the FM phase (top), with corresponding recurrence plots at a recurrence rate $\text{RR}=10\%$ (bottom). When approaching the quantum phase transition at $h=h_c=1$, a qualitative change in the RPs is visible.}
    \label{fig:3regimes}
\end{figure*}

In this work, we use recurrence analysis as a framework to investigate the dynamics of a quantum many-body system. As a principal example, we choose quenches of the one-dimensional transverse-field Ising model (TFIM). The RPs are constructed from evolving two-site correlators of the order parameter, accessible experimentally, and reveal clear visual differences depending on the quenched transverse field strength. We further employ the RQA to accurately recover the (known) critical value of the transverse field, detecting the phase transition in an unsupervised manner from the dynamics, which constitutes a direct application of recurrence analysis in the quantum many-body context.

\textit{Recurrence analysis.}
Given a time series $\lbrace x_1, x_2, \dots, x_T \rbrace$ over $T$ timesteps, where $x_i\in \mathbb{R}$, one can compute the distance matrix $D_{ij}=|x_i-x_j|$ \cite{footnote_multivariate} (see \cite{supplement} for a more general embedding of the signal into a respective $d$-dimensional effective phase space - here we resort to the simplest case of a trivial $d=1$ embedding). The recurrence matrix
\begin{equation}
R_{ij} = \Theta(\varepsilon - D_{ij}),
\end{equation}
where $\Theta(\cdot)$ is the Heaviside step function, captures pairs of times in which the values of the time series are the same within a predefined threshold $\varepsilon$. The recurrence plot \cite{Eckmann1987} maps $R_{ij}=1$ ($R_{ij}=0)$ to black (white) pixels, visually revealing temporal structures and recurring patterns within the signal. To allow for a consistent comparison across different RPs, one can fix the recurrence rate ({RR}): the threshold $\varepsilon$ is set individually for a time series as the {RR-th} percentile of $D_{ij}$ elements except the diagonal.

Beyond the visual inspection of RPs, recurrence quantification analysis provides a set of quantitative measures that characterize the geometrical structures appearing in the RP, primarily based on the statistics of diagonal and vertical line lengths in the recurrence matrix, which reflect different dynamical properties of the system \cite{Webber2005, marwan2007recurrence,MarwanWebber2015}. Here, we focus on four commonly used quantifiers.

Determinism ({DET}) measures the fraction of recurrence points that form diagonal line structures of at least a minimal length $l_{\min}$ (we choose $l_{\min}=2$): ${\mathrm{DET}=\sum_{l=l_{\min}}^{T} l\,P(l) / \sum_{ij} R_{ij}}$, where $P(l)$ denotes the number of diagonal line segments of length $l$ in the RP. Diagonal lines occur when segments of the trajectory evolve in a similar way for several consecutive time steps. Consequently, high DET values indicate a more deterministic or predictable evolution, while low values suggest stochastic or irregular dynamics.

Laminarity ({LAM}) quantifies the fraction of recurrence points that form vertical line structures of length at least $v_{\min}$ (we choose $v_{\min}=2$): ${\mathrm{LAM} =\sum_{v=v_{\min}}^{T} v\,P(v)/\sum_{ij} R_{ij}}$, where $P(v)$ is the number of vertical line segments of length $v$. Vertical structures correspond to intervals during which the system remains close to a particular state. Therefore, large LAM values indicate laminar phases or trapping behavior, where the dynamics evolves slowly or temporarily stabilizes.

Divergence ({DIV}) is defined as $\mathrm{DIV} = 1/L_{\max}$, where $L_{\max}$ denotes the maximum diagonal line length excluding the main diagonal. Since long diagonal lines correspond to trajectories that remain close for extended times, small DIV values indicate stronger predictability. Conversely, large DIV values signal rapid separation of trajectories and are often associated with chaotic dynamics.

The entropy of the diagonal line length distribution ({ENTR}) measures the diversity of predictable time scales present in the dynamics: ${\mathrm{ENTR} = -\sum_{l=l_{\min}}^{T} p(l)\ln p(l)}$, where ${p(l) = P(l) / \sum_{l=l_{\min}}^{T} P(l)}$ is the normalized probability distribution of diagonal line lengths. A higher ENTR indicates a broader distribution of diagonal line lengths, reflecting a more heterogeneous set of predictability times in the dynamics.

\textit{Model.} To benchmark the recurrence analysis toolbox in the quantum context, we consider the transverse-field Ising model in one dimension \cite{pfeuty_ising_1970,calabrese_quantum_2011,calabrese_quantum_2012,heyl_dynamical_2013,dutta_quantum_2015} whose Hamiltonian reads
\begin{equation}
    H(h) = - \sum_{i=1}^L \left[ \sx_i \sx_{i+1}+h \sz_i\right]
\label{eq:H}
\end{equation}
where $\sigma^\alpha_i$ are Pauli matrices at site $i$, and periodic boundary conditions $\sigma^{\alpha}_{L+1}\equiv \sigma^\alpha_1$ are assumed. Apart from the translation invariance, the Hamiltonian has a $\mathbb{Z}_2$ symmetry, corresponding to a rotation of all spins around the $z$ axis by $\pi$, $\sigma^{x,y}_i \rightarrow - \sigma^{x,y}_i$, $\sz_i \rightarrow \sz_i$. At zero temperature, in the thermodynamic limit $L\rightarrow \infty$, the system undergoes a quantum phase transition from the ferromagnetic ({FM}, $h<1$) phase, through the quantum critical point ($h_c=1$), to the paramagnetic ({PM}, $h>1$) phase. In the FM phase the $\mathbb{Z}_2$ symmetry is spontaneously broken and there are two degenerate ground states, with spins collectively aligning in $+x$ or $-x$ directions. In the PM phase, the ground state is nondegenerate, and with increasing field $h$ it becomes closer to a product state of spins aligning in the $+z$ direction. The order parameter of the transition is $\langle \sx_i \rangle$, which becomes non-zero in the FM phase upon the $\mathbb{Z}_2$ symmetry breaking.
The model is exactly solvable by mapping to a quadratic fermionic Hamiltonian \cite{dziarmaga_dynamics_2005,rossini_long_2010, sachdev_quantum_2011, calabrese_quantum_2012}.

Here, we concentrate on quantum quenches from the initial state $\ket{\Psi_0}$, the ground state of $H(h_0)$, realized by an instantaneous change of the transverse field to $h$ and evolution with $H(h)$ for time $t$,
\begin{equation}
    \ket{\Psi_0(t)} = e^{-i H(h) t} \ket{\Psi_0}.
\end{equation}
In the fermionic mapping, the time evolution can be performed exactly at $\mathcal{O}(L^3)$ computational cost using the standard correlation matrix formalism \cite{barouch_statistical_1971,peschel_reduced_2009,calabrese_quantum_2011,sierant_challenges_2022}. 

If the state is prepared on one side of the critical point and evolved on the other, the dynamical quantum phase transition occurring in the system manifests itself as a non-analytical behavior of revival probabilities at equally spaced time intervals \cite{heyl_dynamical_2013}; in contrast, in the present work we only investigate the time dependence of observables.

For simplicity, we consider $h_0\rightarrow \infty$, corresponding to the initial PM state $\ket{\psi_0}=\ket{\uparrow \uparrow\uparrow \dots}$, where ${\sz_i\ket{\uparrow_i}=\ket{\uparrow_i}}$. Because the initial state preserves the $\mathbb{Z}_2$ symmetry, for any $t, h$ the order parameter is equal to zero, $\rho^x(t)=\bra{\Psi_0(t)}\sx_i \ket{\Psi_0(t)}=0$. Thus, to observe the dynamics, we study the two-point correlation function of the order parameter,
\begin{equation}
    \rhoxx(\ell, t) = \bra{\Psi_0(t)}\sx_i\sx_{i+\ell}\ket{\Psi_0(t)}
\end{equation}
where $\ell$ denotes the distance between spins, and which does not depend on the position $i$ due to translational invariance of $H(h)$. An important advantage of $\rhoxx(\ell, t)$, alongside existing theoretical insights into its behavior \cite{rossini_long_2010}, is the possibility to directly measure it experimentally \cite{zhang_observation_2017,lienhard_observing_2018,keesling_quantum_2019,rispoli_quantum_2019,li_probing_2023,leonard_probing_2023}.

The Lieb-Robinson bounds \cite{lieb_finite_1972} establish a minimal time for the correlations to reach distance $\ell$, i.e. a lightcone beyond which $\rhoxx(\ell,t)$ is exponentially small. For a finite field $h$, the maximal propagation velocity of an elementary excitation of the post-quench Hamiltonian is $\vmax=2\min[h,1]$ \cite{calabrese_evolution_2005,calabrese_time_2006,calabrese_quantum_2012}. This defines a natural timescale for $\rhoxx(\ell,t)$, the Fermi time $t_F = \ell/(2\vmax)$, so that $\rhoxx(\ell,t)\approx0$ whenever $t<t_F$.

\textit{Results.} 
We compute the exact time evolution of $\rhoxx(\ell,t)$ in the quench from $h_0\rightarrow \infty$ to different $h=0.1-3$ for a system of size $L=128$ and up to maximal time $t=500$, measuring the correlations at a timestep $\delta t = 0.1$. Inspecting $\rhoxx(\ell, t)$ within the lightcone ($t > t_F$) for a fixed $\ell$, we observe (i) a rich, multiscale structure of oscillations, particularly when close to the critical point, and (ii) qualitative changes of the signals with changing $h$. To visualize and analyze these signals, we employ RPs and RQA.
\begin{figure*}
    \centering
    \includegraphics[width=\linewidth]{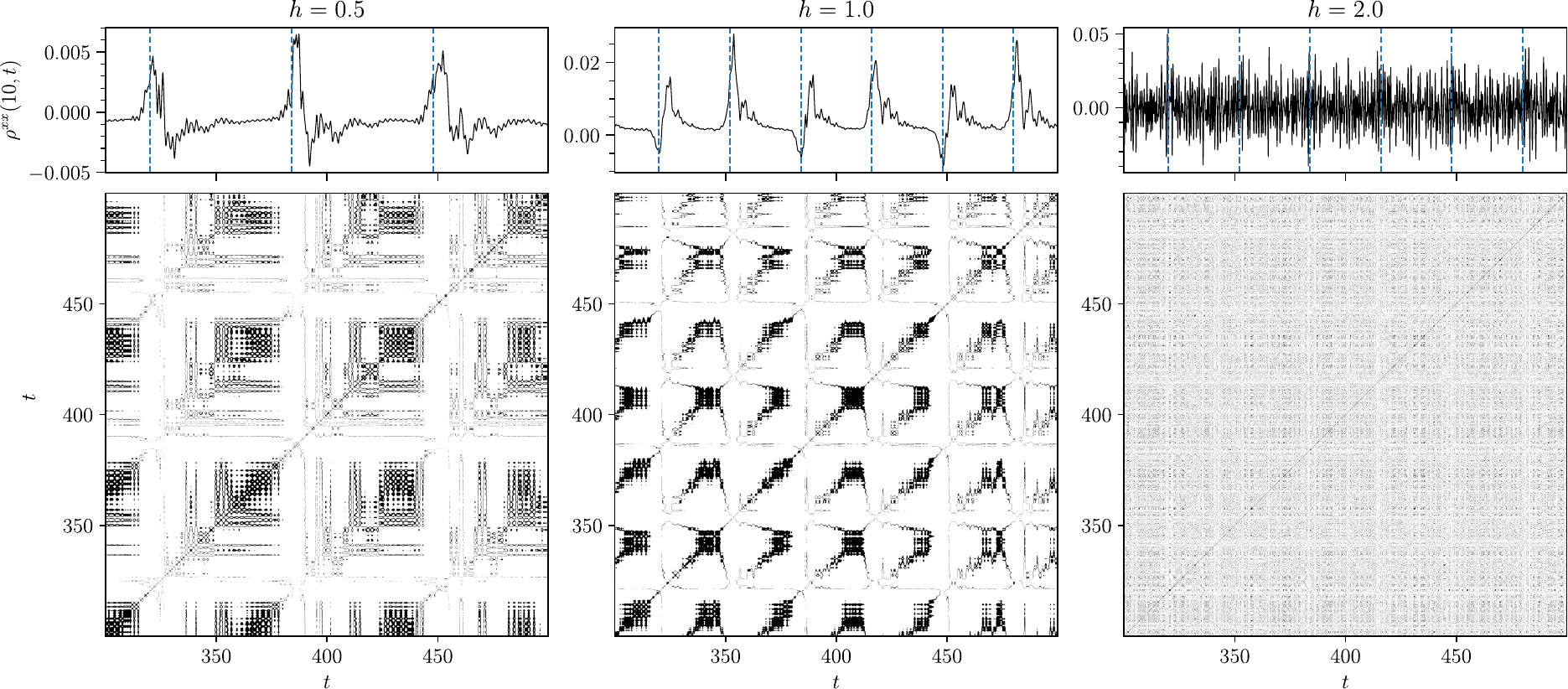}
    \caption{Long-time dynamics of $\rhoxx(10,t)$ correlations in a quench from the paramagnetic state ($h_0 \rightarrow \infty$) to varying $h$ (top), with corresponding recurrence plots, at a recurrence rate $\text{RR}=10\%$ (bottom). When crossing the critical point at $h=h_c=1$, a qualitative change in the RPs is visible. Red dashed lines are plotted for $t = kL/(2\vmax)$, $k\in\mathbb{N}$, i.e. multiples of time it takes for correlations to propagate around the entire system and come back to the initial point.}
    \label{fig:3regimes_long_distance}
\end{figure*}

Figure~\ref{fig:3regimes} (top panels) shows examples of $\rhoxx(1,t)$ between $t=400$ and $t=440$, with a varying distance to the quantum critical point within the FM phase. It is known that this observable is strictly local in the fermionic picture and does not thermalize with time \cite{rossini_effective_2009,rossini_long_2010}. For each of the time series, we present a corresponding RP (bottom panels), fixing the recurrence rate at $\text{RR}=10\%$, aiming at a comparison. We observe a qualitative transition from $h=0.33$, where the repeating signal oscillations give rise to a regular recurrence pattern, to a multi-frequency irregular pattern at the critical point. From the physical point of view, $\rhoxx(1,t)$ is proportional to the density of spin domain walls/kinks \cite{rossini_long_2010}, which are injected in the system by the initial state $\ket{\psi_0}$, not being an eigenstate of $H(h)$, as excitations over the ground state. Thus, the presented RPs characterize the fluctuations of the domain walls density in time. We do not show results for quenches to $h>1$ as the $\rhoxx(1,t)$ observable is exceptional - due to the Kramers-Wannier duality \cite{radicevic_spin_2019} of the Hamiltonian~\eqref{eq:H} under transformation $h \rightarrow 1/h$, one has $\rhoxx(1,t)\vert_h = \rhoxx(1,ht)\vert_{1/h}$. Thus, in the time evolution of $\rhoxx(1,t)$, quenches to the FM and PM phases differ only by a rescaling of the time axis by a constant. While such a change would be clearly visible on the level of recurrence plots, we rather concentrate on less trivial features of the signal.

With $\ell>1$, even though the observable $\rhoxx(\ell, t)$ is a strictly two-body operator in the spin description, it becomes highly non-local in the fermionic degrees of freedom, and is therefore known to thermalize with time despite the model being integrable \cite{rossini_effective_2009, rossini_long_2010}. The long-time dynamics of $\rhoxx(10,t)$ across the entire phase diagram is shown in Fig.~\ref{fig:3regimes_long_distance}. In the top row, the largest peaks result from the finite-size effects and occur around times $t=kL/(2\vmax)$, $k\in \mathbb{N}$, $\vmax=2\min(h,1)$, i.e. multiples of time it takes for the lightcone of correlations to propagate around the entire system and come back to the initial point. Recurrence plots in Fig.~\ref{fig:3regimes_long_distance} capture the corresponding dynamics within and between such cycles. At $h=2.0$ the cycles are much less pronounced. Importantly, the RPs in Fig.~\ref{fig:3regimes_long_distance} distinguish not only the simple changes in the natural timescale caused by $\vmax$ varying with $h$, but also the less trivial recurrence structures, with a stronger qualitative change than in the $\rhoxx(1,t)$ case from Fig.~\ref{fig:3regimes}. In \cite{supplement} we confirm that even after correcting for a multiplicative change in the natural timescale, i.e. taking the Fermi time $t_F$ as a unit of time, the RPs still depict clear differences between quenches to the FM, critical, and PM phases.

\begin{figure*}
    \centering
    \includegraphics[width=\linewidth]{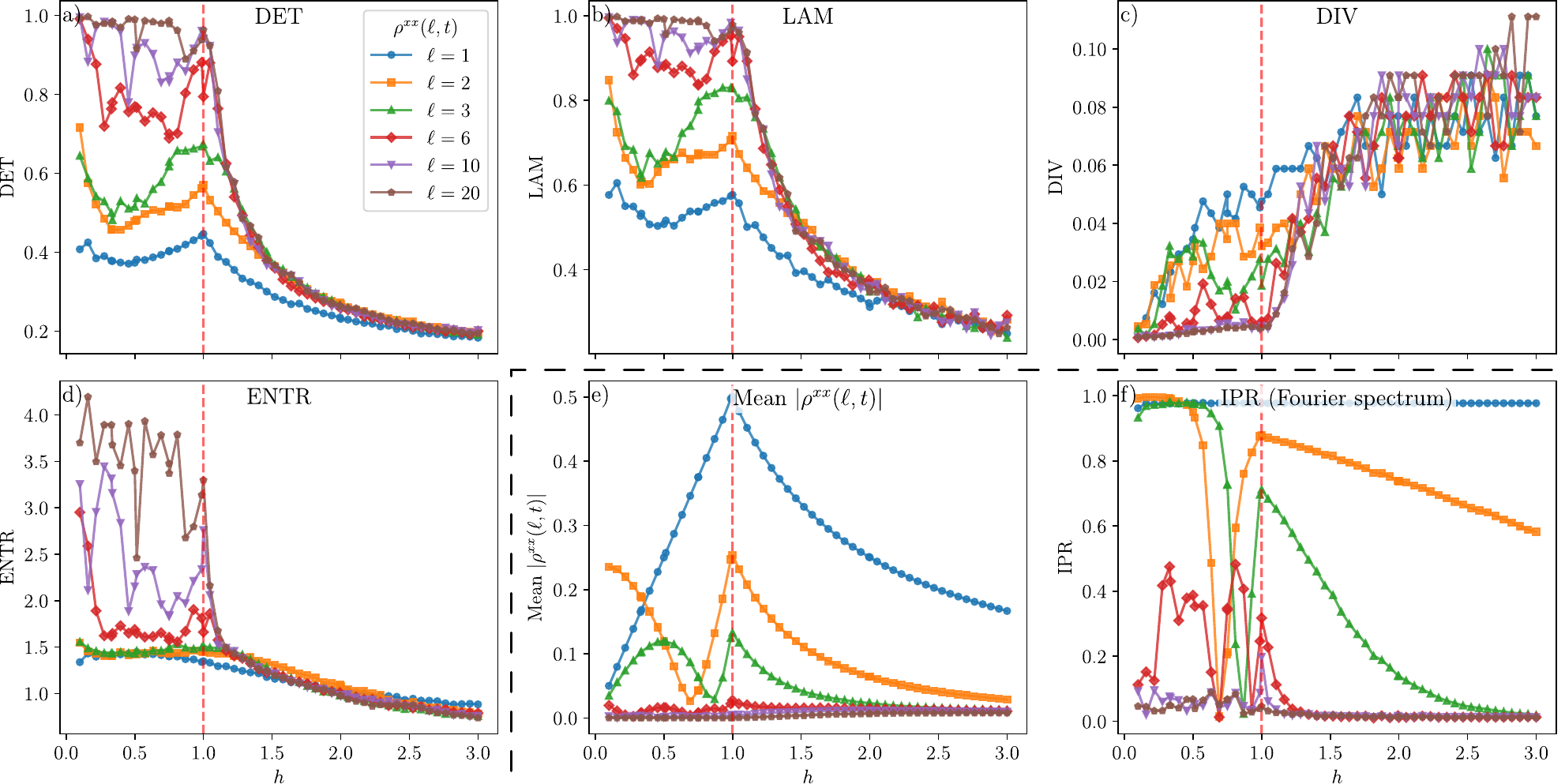}
    \caption{Quantitative analysis of the time evolution of $\rhoxx(\ell, t)$ correlations between $t=300$ and $t=500$. Recurrence quantification analysis (RQA) from the RPs (a-d) reveals a qualitative change at the critical point $h_c=1$ (red dashed line), which becomes more pronounced with an increasing distance $\ell$. Standard measures, such as a mean absolute value of correlations (e) or the inverse participation ratio of the Fourier spectrum (f) also point at the transition, but the trend with $\ell$ is reversed. 
    }
    \label{fig:rqa}
\end{figure*}

To quantify how the RPs vary across the transition, we employ four RQA descriptors and present them as a function of $h$ in Fig.~\ref{fig:rqa}a-d). Quantifiers originate from the RPs at $\text{RR}=10\%$ over the time interval from $t=300$ to $t=500$ (i.e. $2000$ timesteps), and different correlation distances $\ell$. Clearly, a qualitative change at $h_c=1$ is visible in almost all cases. All quantifiers mark the $h > 1$ ($h < 1$) region as "chaotic" ("regular"). 
While for $\ell=1$, the peaked structure in DET and LAM quantifiers results from the self-dual property of $\rhoxx(\ell,t)$ for $h$ and $1/h$, similar peaks are also pronounced for $\ell=2,3$. For $\ell=10,20$, \textit{all} quantifiers correctly recover the transition point, with a plateau for $h<1$ and an abrupt change at $h\geq1$. The DIV and ENTR quantifiers do not reveal a qualitative change for $\ell=1$. We also verified that for the $\ell=1$ curves, the lack of an exact symmetry upon $h \leftrightarrow 1/h$ transformation is caused by i) a multiplicative change of the time axis, to which some quantifiers are susceptible, ii) too early stages of the dynamics considered for the smallest $h\approx 0.1-0.4$.

The most important conclusion from the RQA study is the ability to recover the known critical point at $h=1$ from the numerical analysis of RPs. This constitutes an example of an application of RPs and RQA for an unsupervised quantum phase transition detection \cite{kottmann_unsupervised_2020, lidiak_unsupervised_2020, kaeming_unsupervised_2021, kottmann_unsupervised_2021, kottmann_variational_2021} from the dynamics. Interestingly, the strongest signals pointing at the transition are observed for long ranges $\ell=10,20$ of the two-body correlator $\rhoxx(\ell,t)$. While expected from the comparison of RPs in Figures~\ref{fig:3regimes} and \ref{fig:3regimes_long_distance} due to larger qualitative differences between different $h$ in the latter, such behavior is not observed for more standard time series analysis tools such as in Fig.~\ref{fig:rqa}e-f), not involving any recurrence analysis. Both the mean value of $|\rhoxx(\ell,t)|$ (panel e)) and the inverse participation ratio of the normalized Fourier spectrum of $\rhoxx(\ell,t)$ (panel f), see \cite{supplement} for the definition), give weaker signals with an increasing $\ell$ as probes of the transition.

While our analysis here is limited only to the $\rhoxx(\ell,t)$ observable, we also performed the RQA with the (non-thermalizing) connected correlation function $\rhozzc(\ell,t)$ of the $z$ spin component \cite{rossini_effective_2009, rossini_long_2010}, and could similarly recover the location of the quantum critical point \cite{supplement}, which suggests that the method is applicable to other simple observables.

\textit{Conclusions.} We apply recurrence analysis, a tool well-established in the classical dynamical systems community, to the observables resulting from quantum many-body dynamics. As a prototypical example, we choose quenches of the transverse-field Ising model in one dimension across the phase diagram. Recurrence plots, a pictorial representation of pairs of times at which the time series reaches the same values within a threshold, are used to analyze the evolution of thermalizing or non-thermalizing two-site correlations of the order parameter $\rhoxx(\ell,t)$ after the quench from a deeply paramagnetic state. Qualitative changes of the RPs allow us to distinguish different dynamical regimes depending on the position on the phase diagram after the quench. We further employ several recurrence quantification analysis tools, such as determinism, laminarity, divergence, and entropy, to assess the methods' capabilities to recover the known critical transverse field $h_c=1$. All investigated quantifiers possess a peak or an abrupt drop around $h=h_c$, detecting a qualitative change of the signal, possibly due to the quantum phase transition, in a fully unsupervised manner. Interestingly, the long-range correlators, unlike in standard time- or frequency-space analyses, allow us to recover the critical value of the field with a better resolution than the short-range correlators. Overall, we establish recurrence analysis as a versatile tool for qualitative and quantitative investigations of the dynamics of quantum many-body systems.

\textit{Outlook.} Let us point to possible applications of RPs to the characterization of different dynamical quantum systems. In general, we expect that RPs will be applicable to any non-equilibrium interacting quantum systems which generate signals with a non-trivial structure, somewhere between regular single-frequency oscillations and chaotic, totally noisy patterns. In presence of a sufficiently strong disorder some systems break ergodicity through many-body localization \cite{basko_metal_2006,gornyi_interacting_2005,oganesyan_localization_2007}. RP analysis of the correlation functions across different disorder strengths could supplement current studies \cite{sierant_challenges_2022,sierant_mbl_2025} on determining the critical disorder strength for ergodicity breaking. An interesting avenue would also be to investigate systems hosting quantum many-body scars \cite{bernien_probing_2017,turner_weak_2018,serbyn_quantum_2021}, where for some simple initial states the system shows periodic revivals, while for generic states it thermalizes. Signals from the evolution of scarred states, due to their quasi-oscillatory structure, are supposedly well-suited to be detected and analyzed by RPs.

\textit{Acknowledgements}
M. P. S. acknowledge the funding of the São Paulo Research Foundation (FAPESP), project numbers 2024/22136-6 and  2023/07704-5. We thank P. Sierant for a careful reading of the initial version of the manuscript and valuable suggestions and comments that helped us improve it. We acknowledge the use of \verb|pfapack| \cite{wimmer_efficient_2012}, \verb|quspin| \cite{weinberg_quspin_2017,weinberg_quspin_2019} and \verb|pyuynicorn| \cite{donges_unified_2015} software packages.

\input main.bbl
\appendix

\setcounter{equation}{0}
\setcounter{figure}{0}
\setcounter{table}{0}
\makeatletter
\renewcommand{\theequation}{S\arabic{equation}}
\renewcommand{\thefigure}{S\arabic{figure}}
\renewcommand{\thetable}{S\arabic{table}}
\renewcommand{\bibnumfmt}[1]{[S#1]}

\section*{SUPPLEMENTAL MATERIAL}
\section{Time-delay embedding}
\label{app:embedding}
Since many real-world systems provide only a single observed variable, the underlying phase space dynamics is reconstructed using time-delay embedding, following the idea of Takens' embedding theorem \cite{Takens1981, Noakes1991}. In this approach, a $d$-dimensional state vector is formed by combining time-delayed copies of the time series $\lbrace x_1,\dots, x_T\rbrace$,
\begin{equation}
\mathbf{X}_i = \big(x_i, x_{i+\tau}, \ldots, x_{i+(d-1)\tau}\big) \in \mathbb{R}^{d}.
\end{equation}
where $\tau$ denotes the time delay and $d$ the embedding dimension. Then, the distance matrix reads ${D_{ij}=\|\mathbf{X}_i-\mathbf{X}_j\|}$, where $\|.\|$ represents a suitable norm in the $d$-dimensional embedding space. An appropriate choice of the $\tau$, $d$ parameters allows one to faithfully reconstruct the geometry of the underlying dynamical system \cite{kraemer2021, Marwan2023}.

In this work, however, we focused on the simplest representation and construct recurrence plots directly from the original scalar time series, corresponding to the case $d = 1$ without additional embedding, and chose a fixed recurrence rate RR that defines the distance threshold $\varepsilon$. Other choices for obtaining the threshold are possible in the $d$-dimensional embedding case, see \cite{lakshmibala_dynamics_2022} for details.

\section{IPR of the Fourier spectrum}
\label{app:ipr}
The spectral analysis of $\rho^{xx}(\ell, t)$ is performed as follows. Given a time series sampled at $N$ equally spaced points with time step $\Delta t$, we compute the discrete Fourier transform 

\begin{equation}
    \hat{\rho}^{xx}(\ell, \omega_k) = \sum_{n=0}^{N-1} \rho^{xx}(\ell, t_n) e^{-i\omega_k t_n},
\end{equation}                                                  
where $\omega_k = k \Delta \omega$ and $\Delta\omega = 2\pi/(N\Delta t)$. We then define the normalized spectral amplitude
\begin{equation}
    \tilde{\rho}^{xx}(\ell, \omega) = \frac{|\hat{\rho}^{xx}(\ell, \omega)|}{\sqrt{\displaystyle\sum_k |\hat{\rho}^{xx}(\ell, \omega_k)|^2\Delta\omega}},
\end{equation}
which satisfies $\sum_k |\tilde{\rho}^{xx}(\ell, \omega_k)|^2 \Delta\omega = 1$, so that $p_k \equiv |\tilde{\rho}^{xx}(\ell, \omega_k)|^2\Delta\omega$ defines a normalized probability distribution over frequencies.
The degree of localization of the spectrum is quantified by the inverse participation ratio (IPR),
$$\mathrm{IPR}(\ell) = \sum_k p_k^2 = \sum_k |\tilde{\rho}^{xx}(\ell, \omega_k)|^4 (\Delta\omega)^2.$$
The IPR equals $1/N$ for a spectrally flat (fully delocalized in frequencies) signal and approaches $1$ for a purely monochromatic signal.

\section{Correcting for a change in the timescale}
\begin{figure*}
    \centering
    \includegraphics[width=\linewidth]{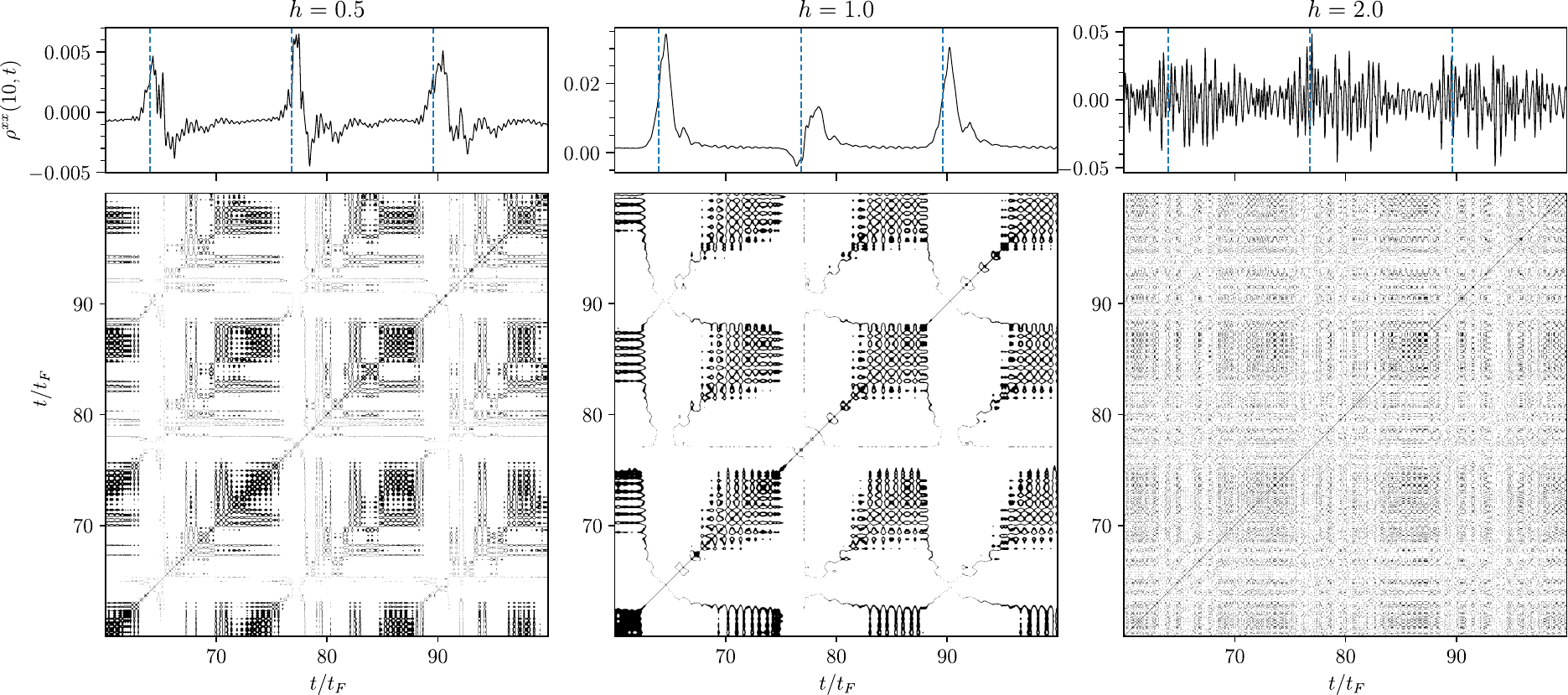}
    \caption{Same as Fig.~\ref{fig:3regimes_long_distance}, but with the time axis rescaled by the Fermi time $t_F$, i.e., corrected by including a theoretically known change in the speed of correlation propagation for varying $h$. Vertical lines correspond to the times at which the correlation lightcone travels around the entire system.}
    \label{fig:3_regimes_long_distance_rescaled}
\end{figure*}

Changing the transverse field $h$ modifies the characteristic timescale $t_F$ of the dynamics. While this is a valid feature to distinguish between different phases, we rather concentrate on distinguishing less trivial structures present in the time series of the observables. Figure~\ref{fig:3_regimes_long_distance_rescaled} shows RPs from Fig.~\ref{fig:3regimes_long_distance} with the time axis given in units of the natural Fermi time $t_F$. Indeed, the largest peaks now have the same interval in time for $h=0.5, 1.0, 2.0$. Clearly, qualitative differences in the RPs are still visible in all three regimes.

\begin{figure*}
\centering
    \includegraphics[width=\linewidth]{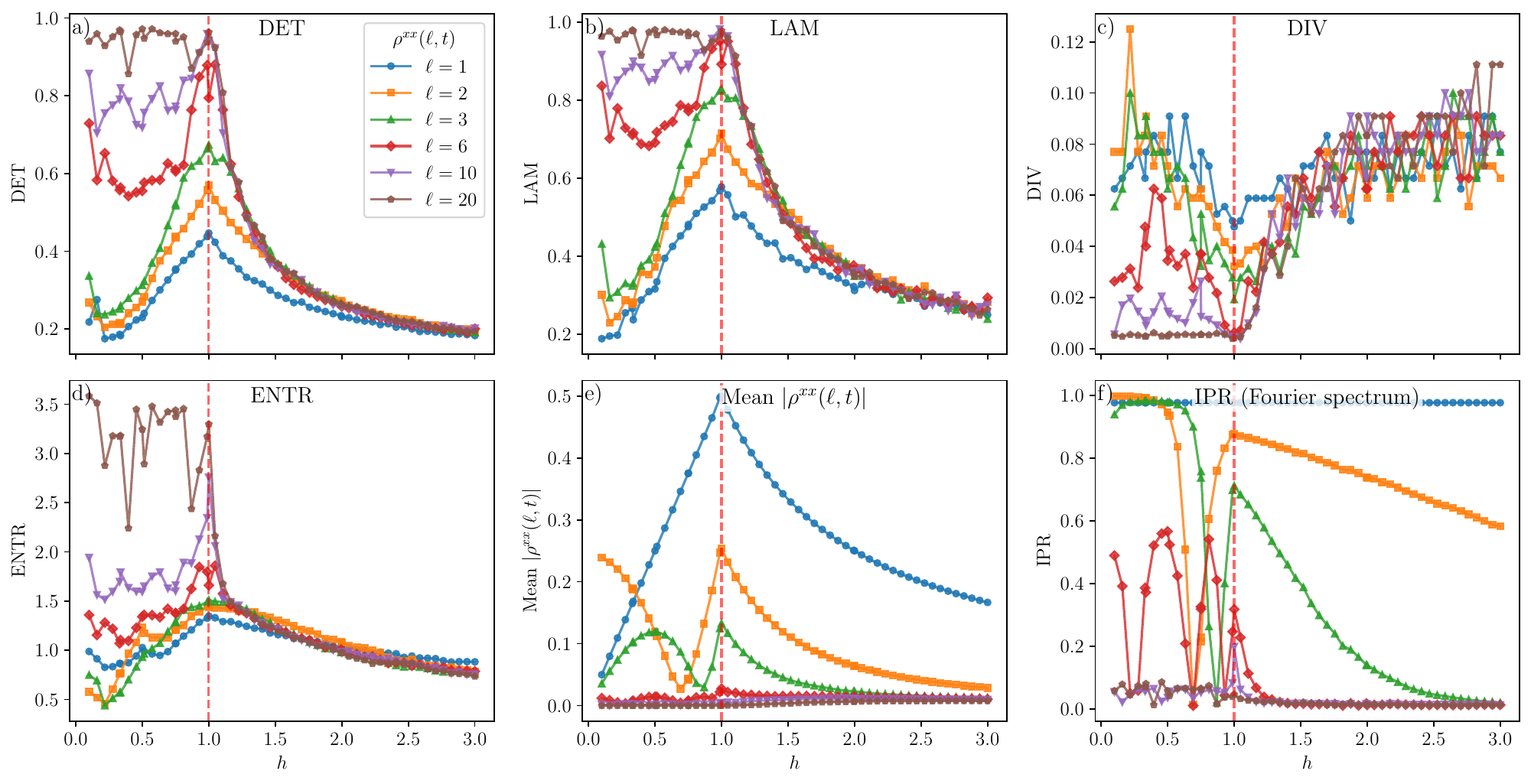}
    \caption{Recurrence quantification analysis, same as in Fig.~\ref{fig:rqa}, but with the time axis rescaled by $t \to t \min(h,1)$ to correct for a trivial change in the timescale with $h$. The RQA still recovers $h=1$ as a point where the temporal behavior of the system changes qualitatively.}
    \label{fig:rqa_rescaled}
    
\end{figure*}

To make sure that our RQA is not merely detecting a trivial change in the timescale, we perform the RQA on signals with time rescaled as $t\rightarrow t\min(h,1)$, and interpolated to zeroth order on the same time grid, $t=300$ to $t=500$, with timestep $\delta t=0.1$. While a physically motivated rescaling would be $t \to t/t_F$, the rescaling factor would then depend on the correlation distance; we chose the simpler rescaling $t\rightarrow t\min(h,1)$ not to obscure the final result by including different numbers of points in time for different $\ell$. Figure~\ref{fig:rqa_rescaled} shows the new RQA of the correlations with rescaled time axis. Peaks and abrupt changes of the quantifiers around $h_c=1$ are now even more strongly pronounced than in the original Fig.~\ref{fig:rqa}, and confirm that non-trivial features of signal, different from just time rescaling, are detected by the RQA.

\section{Recurrence quantification analysis of the ZZ correlations}
\label{app:zz}
\begin{figure*}
    \includegraphics[width=\linewidth]{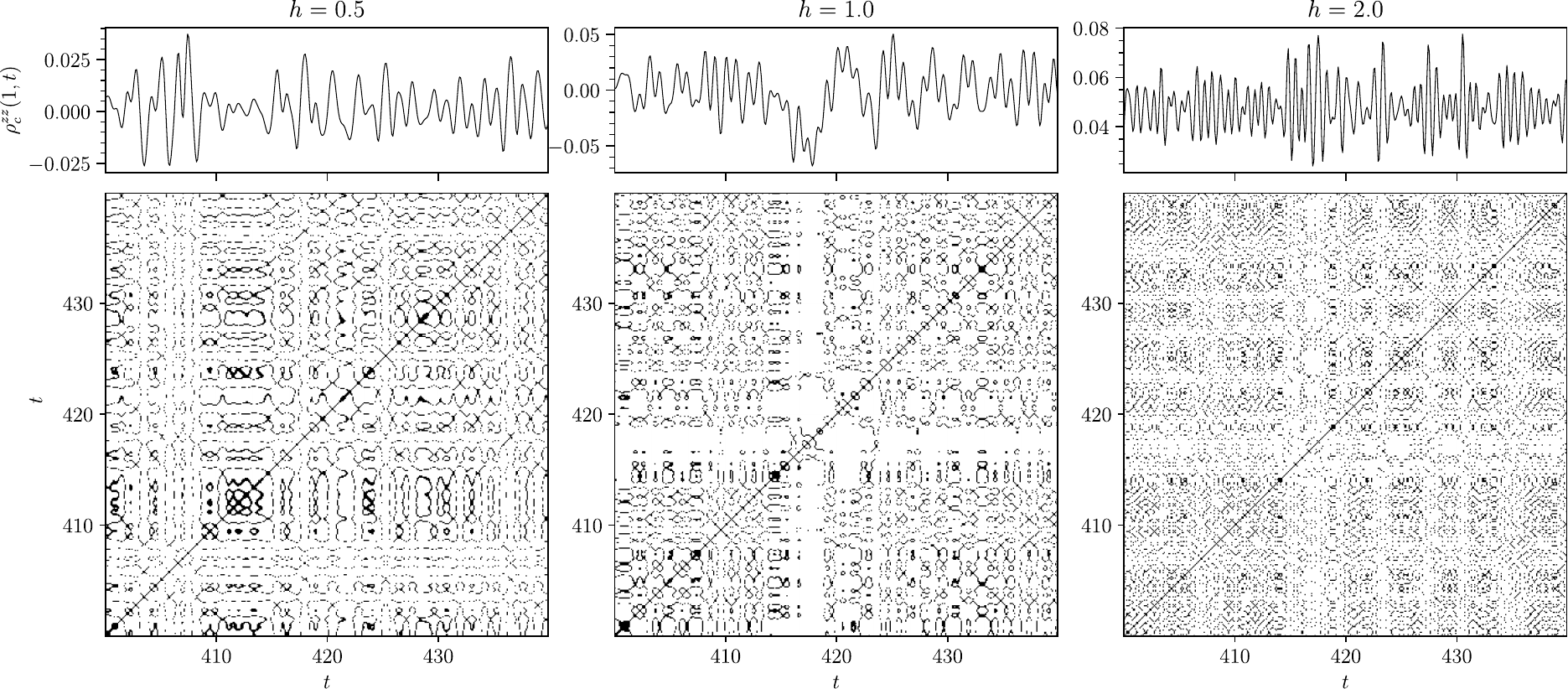}
    \caption{Same as Fig.~\ref{fig:3regimes} but for the connected two-site correlation function $\rho^{zz}_c(1,t)$, which does not involve the order parameter. Qualitative differences between phases are visible.}
    \label{fig:3regimesZZ}
\end{figure*}

\begin{figure*}
    \includegraphics[width=\linewidth]{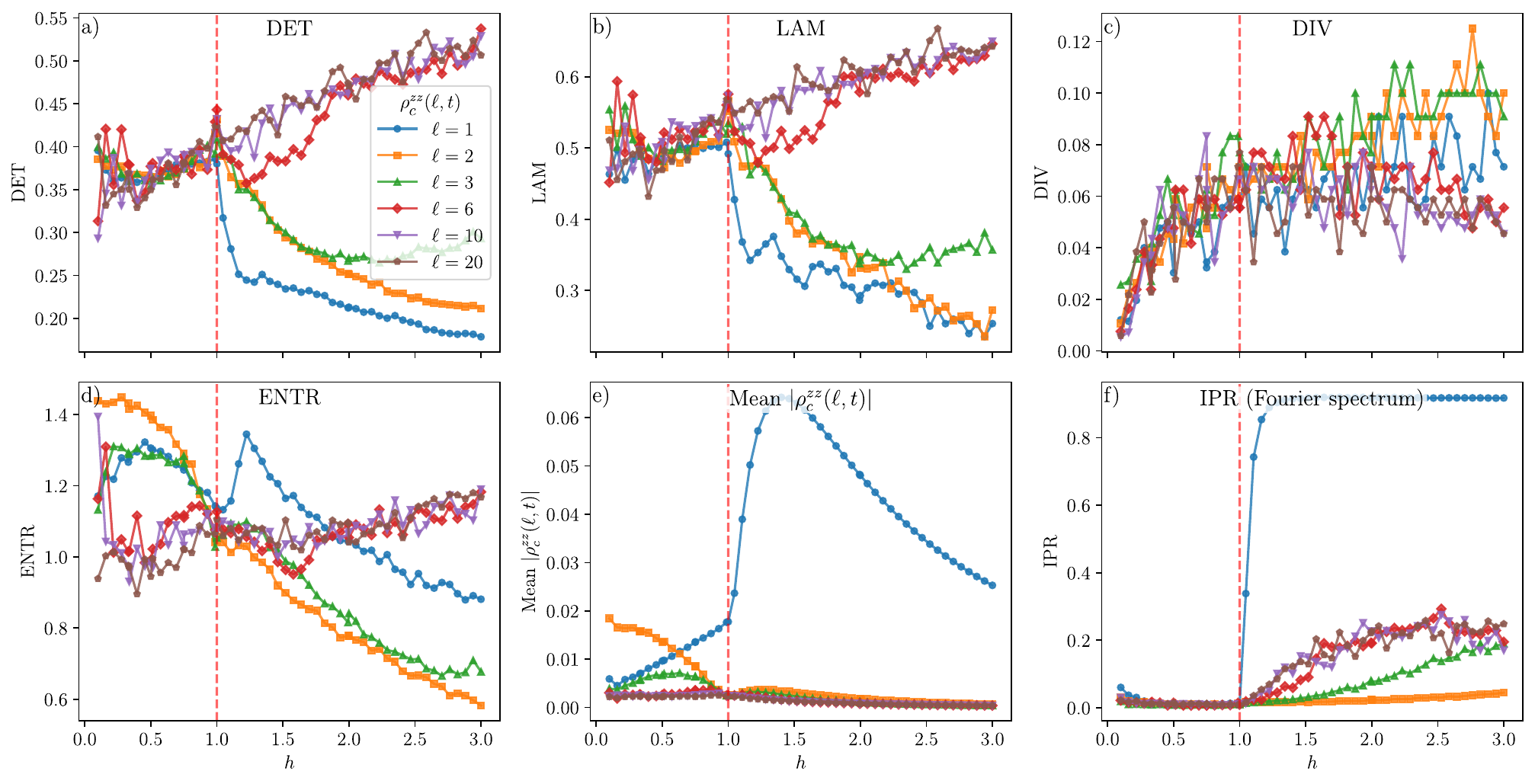}
    \caption{Recurrence quantification analysis for $\rho^{zz}_c(\ell,t)$ correlations between time $t=300$ and $t=500$. Drops or peaks of the quantifiers around $h_c=1$ can be noticed for DET, LAM, and ENTR.}
    \label{fig:rqa_zzc}
\end{figure*}

In addition to the $\rhoxx(\ell,t)$ correlations of the order parameter, in order to verify whether the transition can be detected from other observables, we study the connected correlation function of the spin-$z$ component,
\begin{eqnarray}
    \rhozzc(\ell,t)=\bra{\Psi_0(t)}\sz_i\sz_{i+\ell}\ket{\Psi_0(t)} - (\bra{\Psi_0(t)}\sz_i\ket{\Psi_0(t)})^2.\nonumber
\end{eqnarray}
Exemplary RPs for $\rhozzc(1,t)$ are shown in Fig.~\ref{fig:3regimesZZ} and point to qualitative changes when crossing the critical point.
Results of the RQA for the same parameters as in Fig.~\ref{fig:rqa} of the main text are presented in Fig.~\ref{fig:rqa_zzc}. While the peaked structures around the critical point at $h_c=1$ are generally less pronounced than for the correlators of the order parameter in Fig.~\ref{fig:rqa}, they are present in the DET, LAM and ENTR plots for $\ell\leq 6$. This confirms that a successful unsupervised detection of the phase transition with RQA from $\rhoxx(\ell,t)$ is not feature resulting from a specific choice of the observable, but should apply also for other simple observables. However, here, taking long-range correlations with $\ell=10, 20$ does not allow to detect the transition. We separately verified that rescaling the time axis like in Fig.~\ref{fig:rqa_rescaled} also makes the peaks around $h=1$ more pronounced (plots not shown).

\end{document}

%% file: main.bbl
%

%% file: main.bbl
\begin{thebibliography}{95}
\bibitem [{foo()}]{footnote_multivariate}%
  \BibitemOpen
  \href@noop {} {\bibinfo {title} {Here we consider univariate time series for
  simplicity, but the definitions carry over to the multivariate case by
  replacing the absolute value with a suitable norm.}}\BibitemShut {Stop}%
\end{thebibliography}
